\documentclass[aps,prb,twocolumn,superscriptaddress,longbibliography]{revtex4-2}
\pdfoutput=1

\usepackage[utf8]{inputenc}
\usepackage{graphicx}
\usepackage{physics}
\usepackage{upgreek}
\usepackage{color}
\usepackage{amssymb}
\usepackage{gensymb}
\usepackage{floatrow}

\begin{document}

\title{Anomalous reflection of caustic spin-wave beams}

\author{Franz Vilsmeier}
\email[Contact author: ]{franz.vilsmeier@univie.ac.at}
\affiliation{Department of Physics, School of Natural Sciences, Technical University of Munich, Munich, Germany}
\affiliation{Faculty of Physics, University of Vienna, Vienna, Austria}

\author{Christian Back}
\affiliation{Department of Physics, School of Natural Sciences, Technical University of Munich, Munich, Germany}
\affiliation{Center for Quantum Engineering (ZQE), Technical University of Munich, Munich, Germany}

\date{\today}

\begin{abstract}
Reflection at an interface conserves the wave-vector component parallel to it, which fixes the direction of a reflected plane wave. In anisotropic media, however, beams form at caustic points of the iso-frequency contour, where wave vector and group velocity are strongly non-collinear, and it is unclear which of the two descriptions governs the reflected beam.  We image caustic spin-wave beams reflecting at the edges of a yttrium iron garnet waveguide by time-resolved Kerr microscopy, varying field magnitude and angle. 
For a tilted magnetic field, the reflected beam does not carry the wave vector expected from tangential-wave-vector conservation of the incident carrier. Its direction, wavenumber and wavefront angle instead match the outgoing caustic point. Reflecting every plane-wave component of the measured incident field with its tangential wave vector conserved reproduces the reflected beam for all field angles studied. The components add up in phase at the point where the incident beam hits the edge, which therefore acts as a secondary source of a caustic beam. Both descriptions hold at once, one for each component and the other for the beam as a whole.
\end{abstract}

\maketitle

\section{Introduction}

Spin waves, the collective excitations of ordered magnetic
systems, are promising candidates for low-power information
processing in magnonic devices~\cite{Chumak2015,Serga2010,Barman2021,Chumak2022,Flebus2024,Vilsmeier2026}. Central to the functionality of magnonic circuits is the ability to guide and manipulate spin waves through waveguides, including their reflection at boundaries~\cite{Stigloher2016, Gruszecki2014a, Gruszecki2015, Gruszecki2017,Lock2008}. While the reflection of plane spin waves has been studied extensively and shown to obey an anisotropic form of Snell's law~\cite{Stigloher2016,Vashkovskii1988,Lock2008,Gieniusz2014,Hioki2020}, the behavior of more complex spin-wave excitations at interfaces remains largely unexplored.

Caustic spin-wave beams (CSWBs) represent a distinct type of spin-wave propagation arising from anisotropies in the dispersion relation of in-plane magnetized thin films~\cite{Vashkovskii1988,Veerakumar2006,Schneider2010,Wartelle2023,Demidov2009,Gieniusz2013,Martyshkin2024}. At so-called caustic points on the iso-frequency contour, the group velocity direction $\theta_\mathrm{V}$ (the angle between group velocity and external field) becomes stationary along the contour,
\begin{equation}
	\frac{\mathrm{d}\theta_\mathrm{V}}{\mathrm{d}k}=0 .
	\label{eq:causticcond}
\end{equation}
As a result, a broad range of wave vectors all carry energy in the same direction, forming a well-defined, non-diffracting beam of enhanced amplitude~\cite{Schneider2010,Wartelle2023}. In this beam, the group and phase velocities are strongly non-collinear. Analogous caustic beam phenomena have long been established for e.g. acoustic phonons in crystals~\cite{Taylor1969,Northrop1980}. Spin waves offer a uniquely tunable platform for studying this physics, as the external magnetic field continuously controls the dispersion relation and hence the beam and reflection properties with no direct analogue in other wave systems~\cite{Schneider2010,Wartelle2023,Wagle2026,Heussner2020}. In addition, they can be imaged phase-resolved in real space by time-resolved Kerr microscopy~\cite{Stigloher2016} or Brillouin light scattering~\cite{Serga2006, Schneider2010}.

At boundaries, beams of acoustic phonons~\cite{Northrop1984,Wichard1992,Every1992} and light in photonic crystals~\cite{Yu2004} are reflected anomalously, although each plane-wave component conserves its tangential wave vector. For CSWBs, Schneider et al. observed equal angles of incidence and reflection at a film edge with the field along the edge, and unequal angles with the field tilted~\cite{Schneider2010}. They proposed that the edge acts as a secondary source of caustic beams, without computing the reflected beam. The reflection has not been studied systematically as a function of field magnitude and angle, and the secondary-source picture has not been tested quantitatively.

Here, we study the reflection of CSWBs at the edges of a 40\,$\upmu$m wide yttrium iron garnet (YIG) waveguide by time-resolved magneto-optic Kerr effect (TR-MOKE) microscopy, with field magnitude and angle varied. The reflected beam follows the outgoing caustic point of the iso-frequency contour and not the point with the tangential wave vector of the incident carrier. We reconstruct the reflected beam from the measured amplitude and phase of the incident spin-wave field, with every plane-wave component reflected at conserved tangential wave vector, and reproduce it for all field magnitudes and angles. The reflected beam forms where these components add up in phase, so that the edge indeed acts as a secondary source.

\section{Experiment}

\begin{figure}[t]
	\includegraphics[width=\columnwidth]{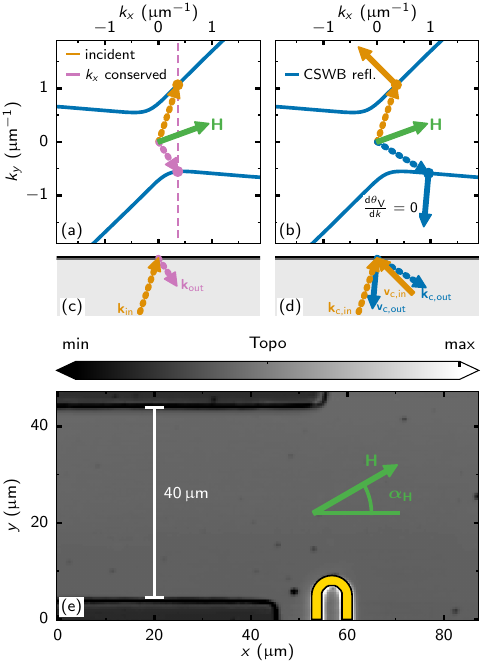}
	\caption{\label{fig:Figure1}
		(a,b)~Iso-frequency contours computed at $f = 1.44$\,GHz, $\mu_0 H = 5$\,mT, and $\alpha_\mathrm{H} = 20^\circ$ for a waveguide edge along $x$. Dotted arrows denote wave vectors $\mathbf{k}$, solid arrows group velocities $\mathbf{v}_\mathrm{g}$, and filled circles mark the caustic points. (a)~Reflection of a single plane-wave component: the reflected wave vector ($k_x$ conserved, pink) lies at the same $k_x$ as the incident one (orange, dashed vertical line) on the outgoing branch ($k_y<0$). (b)~Caustic points of the incident ($k_y>0$, orange) and the reflected beam ($k_y<0$, blue), where $\mathrm{d}\theta_\mathrm{V}/\mathrm{d}k=0$, Eq.~(\ref{eq:causticcond}). For $\alpha_\mathrm{H}\neq 0$, they have different $k_x$. (c,d)~Real-space schematics of the reflection at the waveguide edge (horizontal line): (c)~a single plane-wave component with $\mathbf{k}_\mathrm{in/out}$, (d)~the caustic beam with carrier wave vectors $\mathbf{k}_\mathrm{c,in/out}$ and group velocities $\mathbf{v}_\mathrm{c,in/out}$. (e)~Topographic image of the YIG waveguide recorded simultaneously with magnetization dynamics data shown in Fig.~\ref{fig:Figure2} in the TR-MOKE microscope, with the half-ring antenna (gold) at one end of the 40\,$\upmu$m wide waveguide. The arrow indicates the external magnetic field $\mathbf{H}$ at a representative angle $\alpha_\mathrm{H}$.
	}
\end{figure}

The experiments were conducted on a 200\,nm thick YIG film grown on a gadolinium gallium garnet (GGG) substrate by liquid phase epitaxy. A 40\,$\upmu$m wide magnonic waveguide was fabricated by patterning the YIG film using optical lithography and subsequent argon etching, with the waveguide edges oriented along the $x$-direction. A half-ring-shaped microstrip
antenna (2\,$\upmu$m width) was positioned at one end of the waveguide and driven at a fixed frequency of $f=1.44$\,GHz, exciting spin waves across a broad angular spectrum. An in-plane external magnetic field $\mathbf{H}$ was applied at angle $\alpha_\mathrm{H}$ with respect to the waveguide axis [Fig.~\ref{fig:Figure1}(e)], where $\alpha_\mathrm{H} = 0^\circ$ corresponds to the field aligned with the waveguide edge. The antenna geometry ensures efficient excitation of the caustic points in the low-frequency pocket of the iso-frequency contour, a regime of small carrier wavenumber accessible at moderate applied fields~\cite{Wartelle2023}. One of the two caustic beams emitted symmetrically by the antenna is directed into the waveguide, where it undergoes multiple reflections at the edges, while the other beam propagates freely and serves as a reference.

The magnetization dynamics were imaged using TR-MOKE microscopy~\cite{Dreyer2022,Vilsmeier2024}, which provides spatial maps, referred to as Kerr maps, of the dynamic out-of-plane magnetization component $\delta m_z$ at a fixed phase of the excitation, with a spatial resolution of approximately $0.5\,\upmu$m. Beam parameters such as propagation direction $\theta_\mathrm{V,e}$, carrier wavenumber $k_\mathrm{e}$, and wavefront angle $\varphi_\mathrm{e}$ were extracted from the Kerr maps by two-dimensional least-squares fitting of the beam profiles, following the procedure detailed in the Supplemental Material of Ref.~\cite{Wartelle2023}.

For plane waves, the reflected wave vector follows from the conservation of the wave vector component tangential to the interface ($k_x$) and lies at the same $k_x$ on the outgoing branch of the iso-frequency contour ($k_y<0$)~\cite{Vashkovskii1988,Lock2008,Stigloher2016} [Fig.~\ref{fig:Figure1}(a)]. For CSWBs, wave vector and group velocity are strongly non-collinear, and the beam direction is governed by the stationary group-velocity direction at a caustic point, Eq.~(\ref{eq:causticcond}), rather than by $\mathbf{k}$. The incident beam is carried by the caustic point with $k_y>0$, and the outgoing branch contains the only caustic point accessible to the reflected beam [Fig.~\ref{fig:Figure1}(b)]. If the beam were reflected like a single plane wave, its carrier would instead move to the point with the same $k_x$. Crucially, for $\alpha_\mathrm{H}\neq 0$ the two caustic points have different $k_x$, and the incident and reflected beam directions relative to the edge are unequal [Figs.~\ref{fig:Figure1}(c),(d)]. The field angle rotates the iso-frequency contour and with it the caustic points, thereby steering both the incident and the reflected beam. While every plane-wave component conserves $k_x$, the reflected beam as a whole is governed by the caustic-point condition, and both descriptions hold at the same time. In the following, we examine which carrier the reflected beam follows (Secs.~\ref{sec:magnitude} and~\ref{sec:angle}) and how the beam emerges from the reflected components (Sec.~\ref{sec:reconstruction}).

\begin{figure*}[t]
	\includegraphics[width=\textwidth]{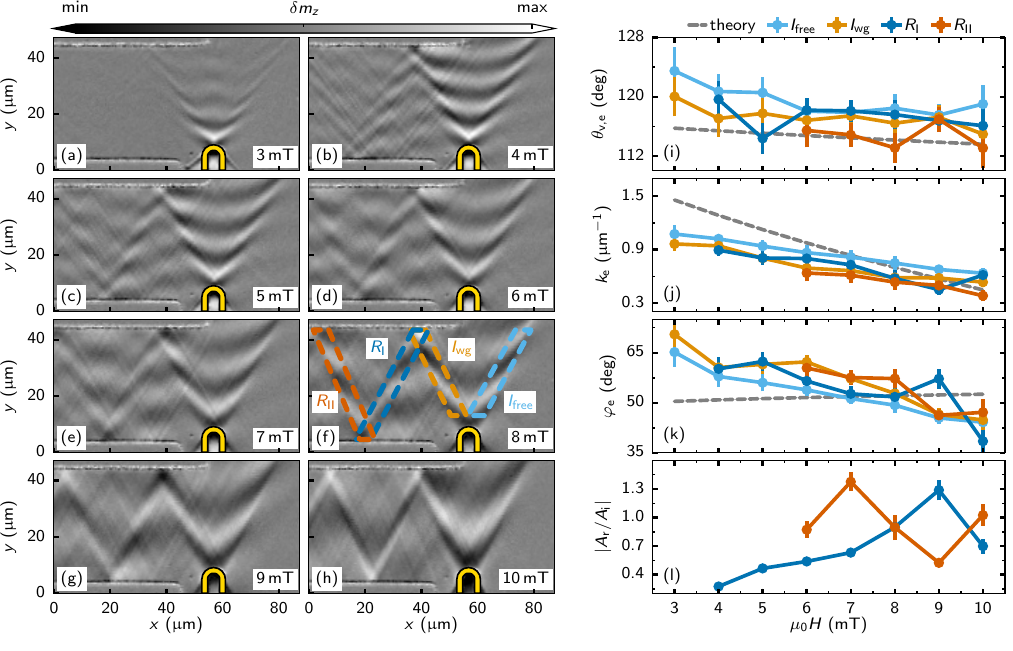}
	\caption{\label{fig:Figure2}
		TR-MOKE maps recorded at $f = 1.44$\,GHz and $\alpha_\mathrm{H} = 0^\circ$ for external magnetic fields $\mu_0 H = 3$--$10$\,mT (a)--(h), all on the same color scale. At 3\,mT (a), no clear reflected beam is observed. At higher fields (b)--(h), successive reflections at the waveguide edges produce a characteristic zig-zag pattern. Dashed contours in (f) indicate the fitted beam extents of the free beam $I_\mathrm{free}$, the incident waveguide beam $I_\mathrm{wg}$, and the reflected beams $R_\mathrm{I}$ and $R_\mathrm{II}$. At high fields (g),(h), the steeper beam angle produces a denser multi-reflection pattern. (i)--(k)~Extracted beam direction $\theta_\mathrm{V,e}$, carrier wavenumber $k_\mathrm{e}$, and wavefront angle $\varphi_\mathrm{e}$ as a function of $\mu_0H$ for $I_\mathrm{free}$, $I_\mathrm{wg}$, $R_\mathrm{I}$, and $R_\mathrm{II}$, together with the predicted caustic-point properties (theory, gray dashed). (l)~Amplitude ratio $|A_\mathrm{r}/A_\mathrm{i}|$ for the first ($R_\mathrm{I}$, blue) and second ($R_\mathrm{II}$, orange) reflection.
	}
\end{figure*}

\section{Field-magnitude dependence}
\label{sec:magnitude}

We first examine the case $\alpha_\mathrm{H} = 0^\circ$, varying the field magnitude to probe different caustic points within the low-frequency pocket~\cite{Wartelle2023}. Each Kerr map shows a pair of beams emitted from the antenna tip, visible as stripe patterns with well-defined propagation directions and tilted wavefronts [Figs.~\ref{fig:Figure2}(a)--(h)]. Their apparent wavelength and direction change as the applied field is increased. At higher fields [Figs.~\ref{fig:Figure2}(b)--(h)], when the beam reaches the first waveguide edge, a reflected beam with an opposite $y$-direction and a wavelength similar to the incoming beam becomes clearly visible, with a mirrored wavefront angle. Subsequent reflections at the opposite edge produce a characteristic zig-zag pattern. The reflected beams maintain the key features of caustic beams: well-defined propagation directions and distinct carrier wavelengths, showing that the caustic nature is preserved upon reflection. The patterned edges also excite secondary caustic features from point-like scattering, contributing to the complex background within the waveguide.

The beam parameters extracted from the Kerr maps are summarized in Figs.~\ref{fig:Figure2}(i)--(k). Above 5\,mT, the beam direction $\theta_\mathrm{V,e}$, carrier wavenumber $k_\mathrm{e}$, and wavefront angle $\varphi_\mathrm{e}$ are in good agreement with the predicted caustic-point properties for the incident beams ($I_\mathrm{free}$, $I_\mathrm{wg}$) and for the reflected beams, even after the second reflection ($R_\mathrm{II}$) [Fig.~\ref{fig:Figure2}(f)]. The reflected beams thus follow the caustic points of the iso-frequency contour.

Below 5\,mT, the measured beam parameters deviate increasingly from the prediction, most clearly in the carrier wavenumber and in the wavefront angle [Figs.~\ref{fig:Figure2}(j),(k)]. We attribute this deviation to the approach to the anticrossing of the fundamental mode with the first perpendicular standing spin wave (PSSW), which forms near 1\,mT at 1.44\,GHz and is neglected in the thin-film model. At 3 mT, the anticrossing is already so close that hybridization in addition strongly damps the spin waves~\cite{Riedel2023,Vilsmeier2024}, and no reflected beam is observed [Fig.~\ref{fig:Figure2}(a)]. The demagnetizing field at the waveguide entrance lowers the effective field further. The beams are also attenuated more strongly at lower fields in general, since already in the thin-film model the group velocity at the caustic points decreases from 2.8\,km/s at 10\,mT to 1.9\,km/s at 3\,mT~\cite{Wartelle2023}.

The amplitude ratio of reflected to incident beam ranges from 0.3 to 1.3 [Fig.~\ref{fig:Figure2}(l)], so that the reflected beam retains a large part of the incident amplitude. Values exceeding unity may arise from background signals excited at the patterned edges, which interfere with the reflected beam within the waveguide. The ratio tends to decrease toward lower fields, consistent with the stronger attenuation discussed above. Reflection losses at the edge may further reduce it. Since these contributions cannot be separated, the ratio serves only qualitative purposes.

\begin{figure*}[t]
	\includegraphics[width=\textwidth]{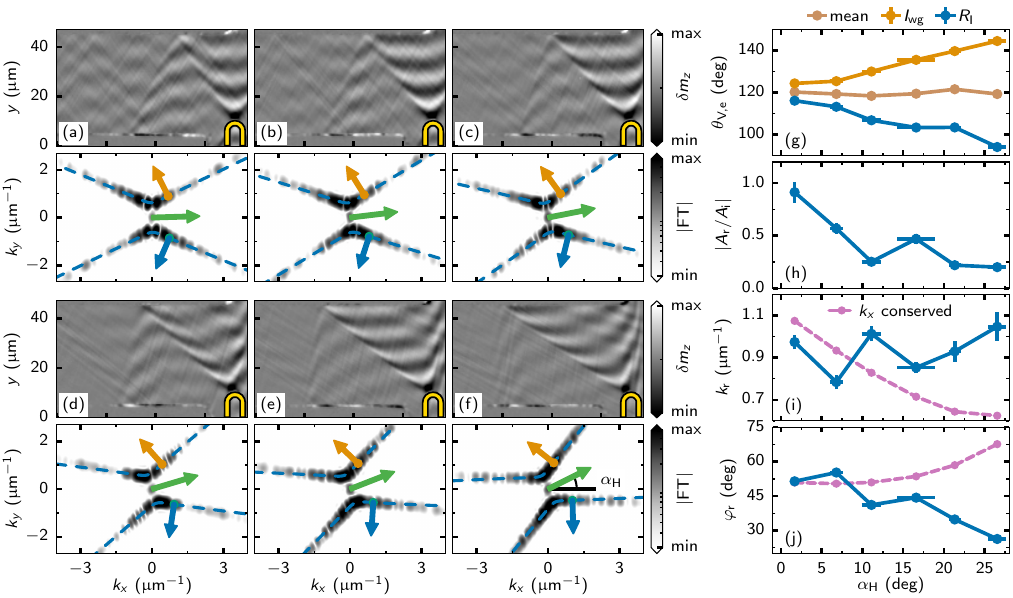}
	\caption{\label{fig:Figure3}
		TR-MOKE maps and reciprocal-space spectra recorded at
		$f = 1.44$\,GHz and $\mu_0 H = 5$\,mT for field angles
		$\alpha_\mathrm{H} \approx 1.7^\circ$, $6.8^\circ$, $11.1^\circ$,
		$16.6^\circ$, $21.3^\circ$, $26.5^\circ$ (a)--(f).
		Below each Kerr map, the corresponding Fourier transform (FT) panel
		shows the measured spin-wave spectrum (greyscale) together with the
		theoretical iso-frequency contour (dashed), the two caustic points
		(colored dots), their group velocity directions (arrows), and the
		external magnetic field direction.
		(g)~Propagation directions $\theta_\mathrm{V,e}$ of the incident beam
		$I_\mathrm{wg}$ and reflected beam $R_\mathrm{I}$ as a function of
		$\alpha_\mathrm{H}$, together with their average. The average remains
		approximately constant, consistent with the $\alpha_\mathrm{H} = 0^\circ$
		caustic point prediction.
		(h)~Amplitude ratio $|A_\mathrm{r}/A_\mathrm{i}|$ as a function of
		$\alpha_\mathrm{H}$.
		(i),(j)~Reflected carrier wavenumber $k_\mathrm{r}$ and wavefront angle $\varphi_\mathrm{r}$ as a function of $\alpha_\mathrm{H}$. Colored symbols are experimental data, and the dashed line is the prediction for the incident caustic point reflected with $k_x$ conserved.
	}
\end{figure*}

\section{Field-angle dependence}
\label{sec:angle}

Having established the field-magnitude dependence, we now vary $\alpha_\mathrm{H}$ at fixed $\mu_0H = 5$\,mT, where the outgoing caustic point and the point with the tangential wave vector of the incident carrier no longer coincide. In each Kerr map [Figs.~\ref{fig:Figure3}(a)--(f)], the rotation of the incident beam direction with $\alpha_\mathrm{H}$ is directly visible, alongside the anomalously reflected beam. As $\alpha_\mathrm{H}$ increases, the incident beam direction rotates correspondingly, while the reflected beam direction rotates in the opposite sense -- a direct consequence of the rotation of the iso-frequency contour. The corresponding Fourier transform spectra below each Kerr map directly reveal the excited spin-wave spectrum, with the theoretical iso-frequency contour, predicted caustic points, and group velocity directions overlaid, showing that the iso-frequency contour rotates with $\alpha_\mathrm{H}$ [Figs.~\ref{fig:Figure3}(a)--(f)]. The predicted group velocity directions are in good agreement with the beam directions observed in the Kerr maps. Quantitative extraction of the beam directions from the Kerr maps confirms that their average remains approximately constant and matches the predicted caustic beam direction for $\alpha_\mathrm{H}=0^\circ$ [Fig.~\ref{fig:Figure3}(g)]. With increasing $\alpha_\mathrm{H}$, the amplitude ratio tends to decrease [Fig.~\ref{fig:Figure3}(h)], which is discussed in Sec.~\ref{sec:reconstruction}.

The reflected carrier wavenumber and wavefront angle show which point the reflected beam follows [Figs.~\ref{fig:Figure3}(i),(j)]. For comparison, we calculate the reflected wave vector for the incident caustic point with $k_x$ conserved on the iso-frequency contour. The measurement and this $k_x$-conserving prediction follow opposite trajectories as $\alpha_\mathrm{H}$ increases. Conservation of $k_x$ gives a monotonically decreasing wavenumber and an increasing wavefront angle, whereas the measured wavenumber shows no clear trend and the wavefront angle decreases. Instead, the measured wavenumber and wavefront angle agree with the outgoing caustic point, whose wavenumber is expected to stay constant and whose wavefront angle decreases as the contour rotates with the field.

\section{Reconstruction of the reflected beam}
\label{sec:reconstruction}

In the field series, the outgoing caustic point coincides with the point of conserved $k_x$. In the field-angle series, the two differ, and the reflected carrier follows the outgoing caustic point. The question arises whether this can still be explained if every plane-wave component is reflected with its tangential wave vector conserved [Fig.~\ref{fig:Figure1}(a)]. To this end, we model the spin-wave field as a superposition of plane waves on the iso-frequency contour.

With the edge along $x$ at $y=y_\mathrm{edge}$, the part of $\delta m_z$ propagating toward the edge is the real part of
\begin{equation}
	\psi_\mathrm{in}(x,y)=\sum_{k_x} S(k_x)\,e^{\,i k_x x+i k_{y,\mathrm{in}}(k_x)\,(y-y_\mathrm{edge})} ,
	\label{eq:in}
\end{equation}
where $S(k_x)$ is the complex amplitude of the component with tangential wave vector $k_x$ at $y=y_\mathrm{edge}$, and $k_{y,\mathrm{in}}>0$ its transverse component on the iso-frequency contour [Fig.~\ref{fig:Figure1}(a)].

For the reflected part, $k_x$ is conserved, so that $k_{y,\mathrm{in}}$ is replaced by $k_{y,\mathrm{out}}<0$ on the outgoing branch of the iso-frequency contour. Each component is reflected with a coefficient $r$, which may in general depend on $k_x$. For simplicity, one $r$ is used for all components, so that the reflected part reads
\begin{equation}
	\psi_\mathrm{r}(x,y)=r\sum_{k_x} S(k_x)\,e^{\,i k_x x+i k_{y,\mathrm{out}}(k_x)\,(y-y_\mathrm{edge})} .
	\label{eq:refl}
\end{equation}
Both $k_{y,\mathrm{in}}(k_x)$ and $k_{y,\mathrm{out}}(k_x)$ are calculated from the iso-frequency contour. The same $S(k_x)$ enters Eqs.~(\ref{eq:in}) and (\ref{eq:refl}). A constant $r$ only scales the reflected part and shifts its phase, so that $r=1$ is used in the reconstruction.

\begin{figure*}[t]
	\centering
	\includegraphics[width=\textwidth]{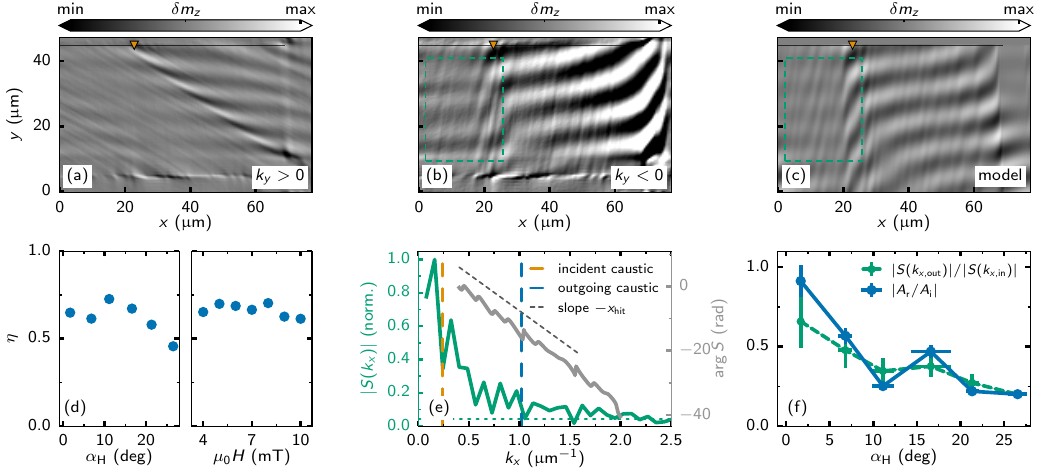}
	\caption{\label{fig:Figure4}
		(a)~Measured incoming ($k_y>0$) and (b)~reflected ($k_y<0$) part, and (c)~model $\mathrm{Re}\,\psi_\mathrm{r}$ multiplied by the best-fitting $r$, at $\mu_0H=5$\,mT and $\alpha_\mathrm{H}=26.5^\circ$. The black line marks the upper edge up to the waveguide entrance, the triangle the hit point, and the dashed box the region of $\eta$. (b) and (c) share one color scale, enhanced by a factor of 3 relative to (a).
		(d)~$\eta$ for the field-angle ($\mu_0H=5$\,mT) and the field series ($\alpha_\mathrm{H}=0^\circ$).
		(e)~Normalized $|S(k_x)|$ at $26.5^\circ$, with the incident and outgoing caustic $k_x$ (vertical lines) and the noise floor (dotted). The right axis shows the phase of $S(k_x)$ above 0.4\,$\upmu$m$^{-1}$, below which a background dominates. The dashed line has the slope $-x_\mathrm{hit}$, with $x_\mathrm{hit}$ the hit position in (a), and is shifted vertically.
		(f)~$|S(k_{x,\mathrm{out}})|/|S(k_{x,\mathrm{in}})|$ with error bars from the noise floor, and the measured amplitude ratio of Fig.~\ref{fig:Figure3}(h).
	}
\end{figure*}

The reflected part is thus emitted from the edge, with the arriving magnetization as its source, as proposed by Schneider et al.~\cite{Schneider2010}. Eq.~(\ref{eq:refl}) is then analogous to the near-field diffraction model of Vlaminck et al.~\cite{Vlaminck2023} and Wagle et al.~\cite{Wagle2026}, with the arriving magnetization in place of the antenna field. As for the beam emitted by an antenna, the reflected components form a beam where their group-velocity direction is stationary, which is discussed after the reconstruction.

To obtain $S(k_x)$ from the measurements, the part of the Kerr map with $k_x>0$ propagating toward the upper edge ($k_y>0$) is extracted by filtering in Fourier space. The part propagating away from the edge ($k_y<0$) serves as the reference for the reconstruction, and both are shown in Figs.~\ref{fig:Figure4}(a),(b) for $\mu_0H=5$\,mT and $\alpha_\mathrm{H}=26.5^\circ$, the case in which the reflected beam deviates most from the one expected from $k_x$ conservation. For this map, the position of the upper edge, determined from the simultaneously recorded topography, is $y_\mathrm{edge}=44.8\,\upmu$m.

$S(k_x)$ is evaluated along the line $y_\mathrm{line}=y_\mathrm{edge}-1.5\,\upmu$m, where the signal is cleaner than directly at the edge. The profile along this line is Fourier transformed in $x$ over the length of the edge up to the waveguide entrance. Each component is then propagated to the edge with its phase factor from Eq.~(\ref{eq:in}),
\begin{equation}
	S(k_x)=\mathcal{F}_x\!\left[\psi_\mathrm{in}(x,y_\mathrm{line})\right]e^{\,ik_{y,\mathrm{in}}(k_x)\,(y_\mathrm{edge}-y_\mathrm{line})} .
	\label{eq:S}
\end{equation}
$S(k_x)$ thus holds the amplitude and phase of every component on arrival at the edge.

\begin{figure*}[t]
	\centering
	\includegraphics[width=\textwidth]{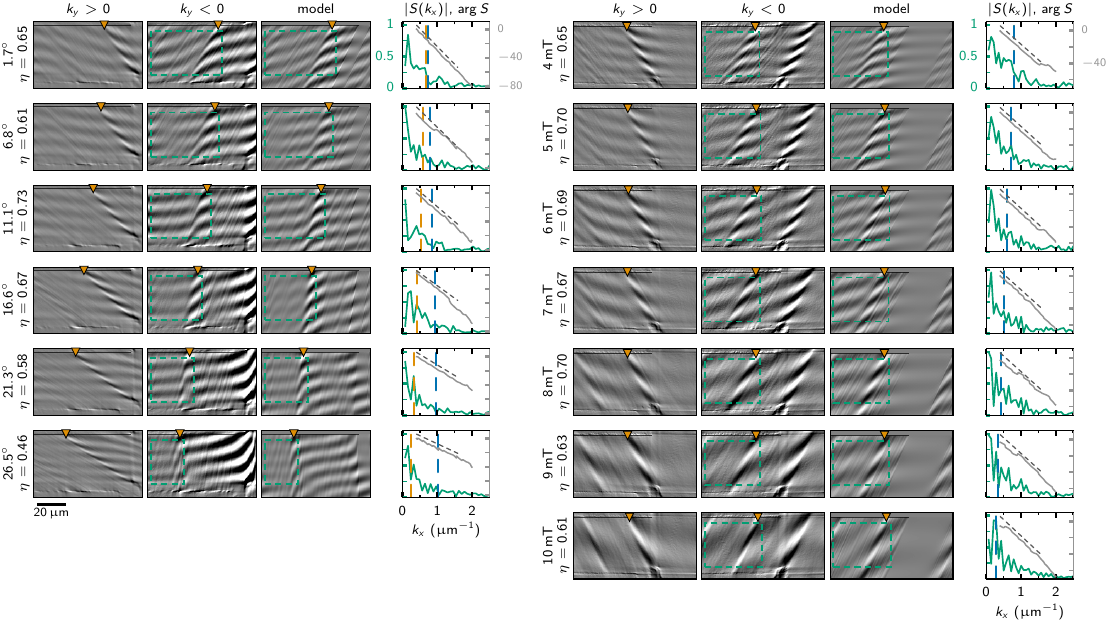}
	\caption{\label{fig:supp}
		Reconstruction for all Kerr maps of the field-angle series at $\mu_0H=5$\,mT (left) and the field series at $\alpha_\mathrm{H}=0^\circ$ (right), one row per map with $\eta$. The columns show the measured incoming ($k_y>0$) and reflected ($k_y<0$) part, the model $\mathrm{Re}\,\psi_\mathrm{r}$ multiplied by the best-fitting $r$, and $|S(k_x)|$ and $\arg S$ as in Fig.~\ref{fig:Figure4}(e), with $|S|$ on one scale per series. In the field series, the incident and outgoing caustic $k_x$ coincide. The reflected part and the model share one color scale per row. Lines, triangles and dashed boxes are as in Fig.~\ref{fig:Figure4}. All maps are shown at the same scale, the scale bar is 20\,$\upmu$m, and the maps of the field-angle series show the field of view of Fig.~\ref{fig:Figure4}(a)--(c).
	}
\end{figure*}

The reflected part is then obtained from Eq.~(\ref{eq:refl}) with $r=1$. The sum is evaluated on the pixel grid of the measured map inside the waveguide and shown in Fig.~\ref{fig:Figure4}(c). The reconstructed map and the measured reflected part agree well [Figs.~\ref{fig:Figure4}(b),(c)]. In particular, the reconstruction forms a beam with a similar position, direction and fringe pattern as the measured one. The same reconstruction was carried out for all Kerr maps and is shown in the Appendix, where the formation of the beam is seen even more clearly [Fig.~\ref{fig:supp}].

To quantify the agreement, we use the squared correlation $\eta$~\cite{Bendat1971} between the reconstructed and the measured map inside the marked region [Figs.~\ref{fig:Figure4}(b),(c)], defined in the Appendix. $\eta$ is the fraction of the measured signal that the model reproduces and does not depend on $r$. It lies between 0.46 and 0.73 for the field-angle series and between 0.61 and 0.70 for the field series [Fig.~\ref{fig:Figure4}(d)]. For comparison, simply reversing $k_y$ of the same components, $k_y\to-k_{y,\mathrm{in}}$, gives $\eta$ below 0.13 from $\alpha_\mathrm{H}=6.8^\circ$ on and 0.004 at $26.5^\circ$, while it coincides with the model for the field series by symmetry.

Why the reflected beam follows the outgoing caustic point is explained by the method of stationary phase. At a point $(x,y)$, only the components of Eq.~(\ref{eq:refl}) add up whose phase $\arg S(k_x)+k_x x+k_{y,\mathrm{out}}(k_x)\,(y-y_\mathrm{edge})$ is stationary in $k_x$. At the edge, this is the case where $x$ equals the negative slope of $\arg S(k_x)$. For the incident beam, this is seen in Fig.~\ref{fig:Figure4}(e), where magnitude and phase of $S(k_x)$ are displayed [for all maps, see Fig.~\ref{fig:supp}]. Its components span a broad range of $k_x$ in which the phase is linear, with a slope that agrees well with $-x_\mathrm{hit}$, where $x_\mathrm{hit}$ is the position at which the beam hits the edge. All of them thus add up at the hit point. The reflected part is excited with the same spectrum and therefore starts there as well. Inside the waveguide, the stationary point of each reflected component moves along its group velocity. A beam forms only where this direction is stationary, Eq.~(\ref{eq:causticcond}), since a broad range of components then shares one line, which is the case at the outgoing caustic point~\cite{Wartelle2023}. The full sum of Eq.~(\ref{eq:refl}) reproduces this beam [Fig.~\ref{fig:Figure4}(c)]. The reflected beam thus forms like the incident one, with the hit point in place of the antenna as its source [Fig.~\ref{fig:Figure1}(d)]~\cite{Schneider2010}.

This can also explain why the reflected beam becomes weaker with $\alpha_\mathrm{H}$, since a caustic beam forms efficiently only if the caustic point lies within the spectrum of its source~\cite{Schneider2010}. At $26.5^\circ$, the outgoing caustic $k_x$ lies in the tail of $|S(k_x)|$, far from the incident one [vertical lines in Fig.~\ref{fig:Figure4}(e)]. This mismatch grows with $\alpha_\mathrm{H}$ [Fig.~\ref{fig:supp}], so that fewer components contribute to the reflected beam. Accordingly, the ratio $|S(k_{x,\mathrm{out}})|/|S(k_{x,\mathrm{in}})|$ decreases with $\alpha_\mathrm{H}$, and the measured amplitude ratio tends to decrease as well [Fig.~\ref{fig:Figure4}(f)]. The comparison remains qualitative, since this ratio equals one for the field series, whereas the measured amplitude ratio varies between 0.3 and 1.3 [Fig.~\ref{fig:Figure2}(l)].

We note that the model treats a single straight edge. It contains neither the opposite edge nor the quantization of $k_y$ in the waveguide, and still reproduces the reflected beam well for all maps [Fig.~\ref{fig:supp}]. The confinement therefore does not appear to influence the formation of the reflected beam strongly. Whether it can matter is seen from the spread of $k_{y,\mathrm{out}}(k_x)$ on the iso-frequency contour over the range in which the phase of $S(k_x)$ is linear, 0.4--1.6\,$\upmu$m$^{-1}$. For the field series, this spread is about 0.5\,$\upmu$m$^{-1}$, six to seven times the mode spacing $\pi/w\approx0.08\,\upmu$m$^{-1}$. In the field-angle series, it decreases with $\alpha_\mathrm{H}$ and falls below the mode spacing from about $20^\circ$ on. From the present maps, however, we cannot make a strong statement about its effect.

\section{Conclusion}

We have studied the reflection of caustic spin-wave beams at the edges of a YIG waveguide as a function of field magnitude and angle. The reflected beam follows the outgoing caustic point of the iso-frequency contour. For a tilted field, its wavefront angle therefore evolves opposite to that of the point with the tangential wave vector of the incident carrier, and its wavenumber does not follow it. A reconstruction from the measured incident spin-wave field, in which every plane-wave component is reflected with its tangential wave vector conserved, reproduces the reflected beam for all field magnitudes and angles. The reflected components start in phase at the hit point and form a beam along the group-velocity direction of the outgoing caustic point, so that the edge acts as a secondary source of caustic beams. The reconstruction uses the measured amplitude and phase of every component and thus shows directly how the reflected beam forms from the incident one. Since this relies only on an anisotropic dispersion with caustic points, it should apply to other anisotropic wave systems. For spin waves, field magnitude and angle tune the direction of the reflected beam.


\begin{acknowledgments}
This work was supported by the Deutsche Forschungsgemeinschaft (DFG, German Research Foundation) under Germany's Excellence Strategy - EXC-2111 - 390814868. Claude Opus 5.5 (Anthropic) was used to assist with editing the manuscript text. All text was reviewed and revised by the authors, who take full responsibility for its content.
\end{acknowledgments}

\appendix*

\section{Reconstruction details}
\label{app:details}

Figure~\ref{fig:supp} shows the reconstruction for all maps, each with a reflected beam of similar position and direction as the measured one. Within each series, the spectra $|S(k_x)|$ along the edge have a similar shape and level. With increasing $\alpha_\mathrm{H}$, only the caustic $k_x$ move: the outgoing one shifts away from the incident one into the tail of $|S(k_x)|$. For most maps, the slope of the phase agrees with $-x_\mathrm{hit}$.

The squared correlation~\cite{Bendat1971} between the reconstructed map $\psi_\mathrm{r}$ and the measured reflected part $\psi_\mathrm{meas}$ is
\begin{equation}
	\eta=\frac{\left|\sum\psi_\mathrm{r}^{*}\,\psi_\mathrm{meas}\right|^{2}}{\sum|\psi_\mathrm{r}|^{2}\sum|\psi_\mathrm{meas}|^{2}} ,
	\label{eq:eta}
\end{equation}
with the sums over the marked region. With the best-fitting $r$, the residual $\sum|\psi_\mathrm{meas}-r\psi_\mathrm{r}|^{2}$ equals $(1-\eta)\sum|\psi_\mathrm{meas}|^{2}$, so that $\eta$ is the fraction of the measured signal that the model reproduces. The region lies left of the hit point, where the measured reflected part contains no free beam, and excludes $4\,\upmu$m next to both edges.

The edge positions were taken from the topography and were not adjusted afterwards. The transform along the edge ends at the waveguide entrance, at $x_\mathrm{end}=69\,\upmu$m for the field-angle series and $56\,\upmu$m for the field series [Fig.~\ref{fig:Figure1}(e)]. Shifting $y_\mathrm{edge}$ by one pixel, $\pm0.5\,\upmu$m, changes $\eta$ by at most 0.1.

The reconstruction and plotting code was written with the assistance of Claude Opus 5.5 (Anthropic), building on the authors' existing analysis routines, and was directed and checked by the authors.

\bibliography{bibliography}

\end{document}